\documentclass[useAMS,usenatbib]{mn2e}

\usepackage{graphics}
\usepackage{epsfig}

\title[Epoch of GC formation]{The SLUGGS survey: inferring the formation epochs of metal-poor and metal-rich globular clusters  
}
\author[D. A. Forbes et al.]{Duncan A. Forbes$^{1}$\thanks{E-mail:
dforbes@swin.edu.au}, Nicola Pastorello$^{1}$, Aaron J. Romanowsky$^{2,3}$, Christopher Usher$^{1}$
\newauthor 
Jean P. Brodie$^{3}$, Jay Strader$^{4}$
\\
$^{1}$Centre for Astrophysics \& Supercomputing, Swinburne
University, Hawthorn VIC 3122, Australia\\
$^{2}$Department of Physics and Astronomy, San Jos\'e State
University, One Washington Square, San Jose, CA 95192, USA\\
$^{3}$University of California Observatories, 1156 High Street, Santa Cruz CA 95064, USA\\
$^{4}$Department of Physics and Astronomy, Michigan State
University, East Lansing, Michigan 48824, USA\\
}
\begin{document}


\pagerange{\pageref{firstpage}--\pageref{lastpage}} \pubyear{2002}

\maketitle

\label{firstpage}

\begin{abstract}

We present a novel, observationally-based framework for the formation
epochs and sites of globular clusters (GCs) in a cosmological
context. Measuring directly the mean ages of the metal-poor and
metal-rich GC subpopulations in our own Galaxy, and in other galaxies, is
observationally challenging. Here we apply an alternative approach  
ulitizing the property that the galaxy mass-metallicity relation is
a strong function of redshift (or look-back age) but is relatively
insensitive to galaxy mass for massive galaxies. Assuming that GCs follow galaxy
mass-metallicity relations that evolve with redshift, one can estimate the mean
formation epochs of the two GC subpopulations by knowing their mean
metallicities and the growth in host galaxy mass
with redshift. 
Recently, the SLUGGS survey has measured the spectroscopic 
metallicities for over 1000 GCs in a dozen massive early-type
galaxies. Here we use these measurements, and our new metallicity matching method, 
to infer a mean age for metal-rich GCs of
11.5 Gyr (z = 2.9) and a range of 12.2 to 12.8 Gyr (4.8 $<$ z $<$ 5.9)
for the metal-poor GCs, depending on whether they 
mostly formed in accreted satellites or in-situ within the main host galaxy.
We compare our values to direct age
measurements for Milky Way GCs and predictions from cosmological models. 
Our findings suggest that reionisation preceded most GC formation, and that it is 
unlikely to be the cause of GC bimodal metallicity distributions.

\end{abstract}


\section{Introduction}

Despite the discovery of multiple stellar populations in Milky Way globular clusters 
(GCs), they are still dominated by a single old age 
population with very little variation in the main sequence turnoff age 
(e.g. Piotto et al. 2007). In this sense they can still be regarded as simple stellar 
populations that  
provide a `fossil record' of 
star cluster formation at the earliest  epochs (see Brodie \& Strader 2006 for a review). 
A key aspect of interpreting this fossil record is the comparison to 
simulations of GC formation within a cosmological context.


Cosmological simulations are now available that model the formation of the two 
subpopulations of GCs commonly seen around large galaxies. Although 
the two subpopulations are observed to have some similar properties (e.g. their mass functions and sizes, to 
first order), they also differ strongly in others (e.g. their metallicity, spatial distribution and kinematics). 
These differences point to different epochs and/or locations for their formation. The two 
subpopulations are commonly denoted by their relative metallicities or colours, i.e. 
metal-poor/blue and metal-rich/red (Peng et al. 2006; Usher et al. 2012). The exact 
division in metallicity varies from galaxy to galaxy. 
Here we refer to two subpopulations as MPGCs and MRGCs.

In this paper we briefly review predictions for the redshift of
formation of these two GC subpopulations in recent cosmological
models.  We also summarise the current situation regarding direct age
measurements of the Milky Way's GC subpopulations, and age estimates
for extragalactic GCs based on integrated spectra.  We then utilise
new spectroscopic metallicities, obtained using the Keck telescope,
for a large sample of GCs from the SAGES Legacy Unifying Globulars and 
GalaxieS (SLUGGS) survey (sluggs.swin.edu.au;
Brodie et al. 2014) and a galaxy mass-metallicity relation that
evolves with redshift to estimate the mean formation epoch of GCs in a
sample of massive early-type galaxies. This method provides an
alternative approach to estimating the mean ages of the two GC
subpopulations. We take advantage of the fact that large numbers of GC
metallicities from integrated spectra are now available from the
SLUGGS survey (Usher et al. 2012, 2015), and that the mass-metallicity
relation is a strong function of redshift (i.e. formation epoch) but
is rather insensitive to mass for massive galaxies.


\section{Cosmological Model Predictions for Globular Cluster Ages}

One of the first cosmological models to predict the formation epoch of the two GC subpopulations was 
Beasley et al. (2002). Their semi-analytical model formed MPGCs in
low mass galaxies and MRGCs in a later gaseous phase but they needed to
truncate the formation of MPGCs at z $>$ 5 in order to reproduce the 
(bimodal) metallicity distribution observed in present day GC systems. Cosmic reionisation was 
identified as the possible process that would interrupt GC formation. 
Using Bennett et al. (2014) cosmological parameters, 
their model predicts 
mean ages of 12.7 Gyr and 10.2 Gyr for the MPGC and MRGCs 
respectively in high mass ellipticals (with slightly younger ages for the MRGCs 
in lower mass host galaxies and/or low density environments). 

In the following discussion we assume the redshift as given in the original work 
and quote a look-back 
time assuming Bennett et al. (2014) 
cosmology (i.e. H$_0$ = 69.6, $\Omega_M$ = 0.286, $\Omega_{vac}$ = 0.714). 
We use the online calculator of Wright (2006) which gives the age of a flat Universe in this 
cosmology of 13.72 Gyr.

In the Santos (2003) model, MPGCs also form prior 
to reionisation at z $\ge$ 7 (ages $\ge$ 13.0 Gyr) 
with reionisation expicitly suppressing any further GC 
formation until it ended. After $\le$ 1.5 Gyrs, the next generation of 
GCs that form were metal enriched. 

In the cosmological simulations of Bekki et al. (2008), GCs form in
strong star formation episodes over a large range of redshift. The
mean formation epoch of MPGCs is z = 5.7 (12.7 Gyr) and z= 4.3 (12.3 Gyr) for MRGCs.

Griffen et al. (2010) used the Aquarius simulation to resolve the dark
matter minihalos which they associated with MPGC formation. In their
model, the formation of MPGC was truncated by ionisation from the
first-formed GCs. They predict MPGC formation to begin around z = 22
(13.6 Gyr) and end by z = 13 (13.4 Gyr). Their favoured formation for
MRGCs is in gaseous major mergers, which give rise to a broad range of
ages, i.e.  7--13.3 Gyrs.

Tonini (2013) modelled the hierarchical assembly of galaxies using a Monte Carlo technique. 
In her model MPGCs form in low
mass satellites at z $\sim$ 3--4 (11.5--12.2 Gyr), with MRGCs forming later in a dissipative
phase within the main galaxy at z $\sim$ 2 (10.4 Gyr). Both GC subpopulations
are imprinted with a metallicity given by 
the galaxy mass-metallicity relation at that redshift. The satellites
are accreted between redshift 4 and 0, with the accretion rate based on the 
Millennium cosmological simulation (Springel et al. 2005). 
If the satellite has MRGCs,
they are still relatively metal-poor compared to the larger main galaxy. In
the case of gas-rich satellites, some new GCs may form and give rise to
an intermediate metallicity GC subpopulation. However, observationally
such intermediate metallicity peaks in well-studied GC systems are
quite rare (Peng et al. 2006).

Based on the earlier cosmological model of Muratov \& Gnedin (2010),
Li \& Gnedin (2014) focused on modelling the GC systems of early-type
galaxies in the Virgo cluster.  In their model GCs form in gas-rich mergers, with early (minor) mergers 
leading to mostly MPGCs and later (major) mergers favouring the formation of MRGCs. Like 
Tonini (2013), the resulting GC metallicities are driven by an evolving galaxy mass-metallicity relation.
Thus a single mechanism produces differences in the GC metallicity and age distributions.
They predict MPGCs to form at redshifts z = 4--6 (12.2--12.8 Gyr) with 
MRGCs peaking around z
= 3--4 but continuing until z $\sim$ 1 (7.8 Gyr). 
Thus the MRGCs could be from zero up to 5 Gyr
younger than their MPGC counterparts.

Building on the method of Moore et al. (2006) and Spitler et
al. (2012), Corbett Moran et al. (2014) associated MPGCs with rare,
overdense peaks which collapse early in the Universe. In particular,
they modelled the formation of MPGCs in a Virgo cluster like
environment, finding a formation redshift that best matches the
observed distribution of MPGCs around M87 of z $\sim$ 9 (13.1 
Gyr). Like several other studies, they invoked reionisation to
ultimately truncate MPGC formation.

Trenti, Padoan \& Jiminez (2015) have proposed that MPGCs form 
in the merger of gas-rich dark matter minihalos (similar to earlier works of 
Bromm \& Clarke 2002 and Boley et al. 2009). 
These minihalos are later stripped of their dark 
matter via tidal interactions within the main galaxy halo. 
They predict a mean formation redshift for MPGCs of z = 9.3 (13.2 Gyr).  
MRGCs are formed later in mergers between 
minihalos but a prediction for their mean age is not given. 

Focusing on a Milky-Way type galaxy in the Via Lactea II simulation, 
Katz \& Ricotti (2014) found that the MPGCs are
dominated by those accreted from satellites but there is also a
contribution from MPGCs formed in-situ. The reverse is true of MRGCs,
which mostly form in-situ but have an accreted component. In the
case of the Milky Way, they predict that over half of the existing GC system
was accreted. Their favoured model predicts MPGC formation 
to occur at z = 7--12 (13.0--13.4 Gyr) and 
z $\sim$ 2 (10.4 Gyr) for the MRGCs, with the bulk of GC 
accretion at redshifts less than 4.

As discussed above, the various cosmological models make 
predictions for the 
formation epoch and therefore the ages of the MPGC and MRGC subpopulations. 
Next we briefly review the observational studies of GC ages.




\section{Observational constraints on globular cluster ages}

Determining the absolute age of Milky Way GCs from their colour magnitude diagrams (CMDs) 
has proved problematic 
over the years due to uncertainties in stellar evolutionary models, intrinsic 
abundance variations, foreground dust corrections, helium content 
and the assumed distance. Nevertheless, 
studies fitting the main sequence turnoff (MSTO) in deep HST CMDs 
have converged on an age for the oldest MPGCs of 12.5 Gyr (VandenBerg et al. 2013) to 
12.8 Gyr (Marin-Franch et al. 2009).
Excluding the clearly younger MRGCs that have a different age-metallicity relation and 
can be associated with 
accretion events (Forbes \& Bridges 2010), 
the MRGCs are either roughly coeval with the MPGCs (Marin-Franch 
et al. 2009) or systematically younger by 1.5 Gyr with a mean age of 11.0 Gyr
(Vandenbergh  et al. (2013) .

Measuring the age of Milky Way GCs from the luminosity fading (i.e. cooling) sequence of 
white dwarfs offers 
an alternative method of obtaining absolute ages that is less sensitive to metallicity 
than MSTO-based methods. 
To date less than half a dozen GCs have published white dwarf cooling track ages. 
They include the metal-poor GC M4 for which Bedin et al. (2009) measured an age of 11.6 Gyr. They estimated a measurement error of 
$\pm$ 0.6 Gyr but noted that the model uncertainties could be as high as $\pm$ 2 Gyr. 
More recently, Hansen et al. (2013) applied this method to two other Milky Way GCs.
They found 
a relatively young age of 9.9 $\pm$ 0.7 Gyr for the metal-rich GC 47 Tuc (NGC 104), which 
is 2.0 $\pm$ 0.5 Gyr younger than their metal-poor GC NGC 6397.
However, also using the cooling sequence method but with different models, Garcia-Berro et 
al. (2014) found an age of 12 Gyr with an uncertainty of $\le$ 1 Gyr for 47 Tuc. Thus the two white 
dwarf cooling ages for this GC are mutually inconsistent. 
Using an updated metallicity, Hansen et al. noted that the eclipsing binary V69 in 47 Tuc 
has an age of 10.39 $\pm$ 0.54 Gyr -- this age and uncertainty being consistent with both the 
Hansen et al. and Garcia-Berro et al. values. Clearly a larger sample of bulge and halo GCs need to be 
studied in this way, and modelling differences reduced, before we can make robust general 
conclusions about the absolute (or relative) ages of the two subpopulations of GCs in the 
Milky Way using white dwarf cooling tracks.



Estimating the relative ages of GCs beyond the Local Group requires
measurements of the integrated light, with the best method being an
analysis of high signal-to-noise optical spectra. The situation for
extragalactic GCs is summarised in the review by Brodie \& Strader
(2006), i.e. the bulk of age-dated GCs are very old ($>$10 Gyr) with
only a small fraction of those observed having young or intermdeiate
ages. In the meta-analysis of Strader et al. (2005) it was concluded
that the best Keck spectra available could not separate the mean ages
of the metal-poor and metal-rich GC subpopulations to better than 1-2
Gyrs. Using spectra from the VLT, Puzia et al. (2005) found hints of a
slightly younger mean age for metal-rich GCs but noted that the
significance of the result needed verification from larger samples.
Spectroscopic studies of GCs tend to be dominated by GCs  
located in the inner regions of galaxies. 
The situation regarding the ages of GC subpopulations
beyond the Local Group is largely unchanged since 2005.

\section{Using the mass-metallicity relation to constrain globular cluster ages}

An alternative approach to directly measuring GC ages is to match GC metallicities to 
the galaxy mass-metallicity relation at different redshifts and hence infer the GC 
epoch of formation. A key assumption is that the mean metallicity of a GC
subpopulation is determined by the same gas that forms the stars within a 
galaxy. Indeed, we know that a significant fraction of galaxy stars
		orginally form in star clusters, which are later
		disrupted (Lada \& Lada 2003; Bastian et al. 2013). 
We can then use the redshift evolution of the galaxy
		mass-metallicity
		relation to estimate the formation epoch (and hence their 
		look-back age, thanks to the era of precision
		cosmology) of MRGC and MPGCs.
This method has the advantage of using
		easier to obtain mean metallicities of extragalactic GCs
		rather than very challenging direct measurements of
		their age.

Spitler (2010) used this approach to focus on the mean metallicity of MRGCs compared to
those of the host galaxy field stars as a function of galaxy stellar
mass (finding a redshift for MRGC formation of z $\sim$ 3.5). Age limits for both MPGC and MRGCs in Virgo galaxies were investigated using a 
similar method by Spitler et al. (2012). This work gave wide ranges for the epochs of GC 
formation, i.e. 2 $<$ z $<$ 4 for MRGCs and 7 $<$ z $<$ 10 for MPGCs.  Spitler et al. concluded that MPGCs formed well within the epoch of reionisation. 
However, a major limitation of these works is that the GC metallicities
were based on observed broad-band colours under the assumption of old
ages, and the GC colour to metallicity transformation is known to 
vary on a galaxy-to-galaxy basis 
(Usher et al. 2015). 
If the mean ages are younger than assumed by Spitler then the redshift of
formation found by Spitler is an upper limit. Shapiro et al. (2010)
employed a similar approach to show that the mean (photometric)
metallicity of MRGCs is consistent with the 
mass-metallicity relation for z $\sim$ 2 star forming turbulent disks.

A better approach is to use mean GC metallicities based on
spectroscopy. Until recently this was only possible for small samples
of GCs and galaxies. 
However, Usher et al. (2012, 2015) and Pastorello et al. (2015) have presented
spectroscopic metallicities for over 1000 individual extragalactic
GCs based on measurements of the Calcium Triplet (CaT) lines from the SLUGGS survey (Brodie et al. 2014). Here we use these measurements 
to calculate the mean metallicities of the MPGC and MRGC subpopulations for 11 massive (log M$_{\ast} > 10.5$ M$_{\odot}$) galaxies. 
These are then compared to evolving galaxy mass-stellar metallicity relations to determine the 
associated redshift and hence the mean formation epoch of the GC subpopulations. 
We utilise the fact that the relation is a strong function of redshift but 
depends only weakly on stellar mass for massive galaxies. 



In order to define the mean metallicities of the two GC subpopulations
we start with photometry, which has the
advantage of being available for large numbers of GCs in a given galaxy. 
The photometry of several GC systems were fit using Gaussian
Mixture Modelling (GMM; Muratov \& Gnedin (2010) to separate
them into two subpopulations in (g--i) colour (see Usher et al. 2012
for details). For the subsample of blue and red GCs with 
spectra and sufficient S/N,
individual GC metallicities are obtained from the CaT absorption lines
using the method outlined in Usher et al. (2012), i.e. CaT indices 
are transformed into total [Z/H] metallicities using the single 
stellar population models of Vazdekis et al. (2003). They show good 
agreement with metallicities from the literature with an rms 
scatter of $\le$ 0.2 dex. 
Here we determine the
weighted mean [Z/H] metallicities for these blue and red GC
subpopulations of 11 early-type galaxies. 

Blue GCs are known to reveal a `blue tilt'
(i.e. becoming more metal-rich with increasing luminosity) for masses
above a few 10$^6$ M$_{\odot}$ or M$_i < -11.5$. The tilt is generally
interpreted as due to self-enrichment (Strader \& Smith 2008; Bailin
\& Harris 2009). Here we 
only include those blue GCs less luminous than M$_i = -11$. 
This exercise removes zero to half a
dozen GCs per galaxy and changes the mean metallicity in most cases by
less than 0.1 dex. There is no evidence for a red tilt, so for the red
GCs we make no such correction. The GC data span a large range in
galactocentric effective radii (i.e. from $R\approx0.4~\rm{R_{e}}$ to
$R\approx18~\rm{R_{e}}$) but show only slight negative radial
gradients (Pastorello et al. 2015) so we make no correction for radial
gradients in GC metallicities.

It is possible that our GC mean metallicities could be impacted by biases in our spectroscopic selection. 
Since GC selection is based on the identification of the 
metallicity-dependent CaT feature, low-metallicity GCs may be excluded if the 
CaT absorption lines in their spectra are very weak. The spectroscopic subsample 
also has a bias to more luminous GCs than average.
However, Pastorello et al. (2015) showed that 
spectroscopically- and photometrically-identified blue GCs have similar mean metallicities in 
the five galaxies for which comparable data were available. 
For the MRGCs, Pastorello et al. found a systematic tendency for the 
spectroscopic mean metallicity to be 
lower than the photometric one in a couple of galaxies 
although this difference is within the error on the mean. 

For each galaxy the total stellar mass is taken from Pastorello et al. (2015), which is based on their $K$-band extinction corrected 
magnitude from the 2MASS source catalog (Jarrett et al. 2000).

Although the metallicities for the MPGCs and MRGCs of the Milky Way galaxy are typically 
determined from CMDs, it is interesting to compare them with those for our 11 
massive early-type galaxies from integrated spectra. 
We take the iron abundances ($[Fe/H]$) for 152  Milky Way GCs 
from the 2010 edition of the Harris (1996) catalogue. For the 45 GCs with 
$\alpha$-element abundances ($[\alpha/Fe]$) from Pritzl et al. (2005) we use their 
quoted value, otherwise we assume a mean value, i.e. $[\alpha/Fe]$ = 0.3. The error on this 
mean value from the 45 GCs in Pritzl et al. is less than $\pm$0.1 dex. 
The total metallicities $[Z/H]$ are then 
obtained from equation 4 of Thomas, Maraston \& Bender (2003):
\begin{eqnarray}
[Z/H] = [Fe/H] + 0.94[\alpha/Fe] 
\end{eqnarray}

Using GMM we find that the metallicity distribution of the Milky Way
GCs is highly bimodal, with peaks at $[Z/H]=-1.24 \pm 0.04$ and $-0.24
\pm 0.03$ dex for the MPGCs and MRGCs, respectively. The uncertainty
in the peak values also comes from the GMM fit. For the total stellar
mass of the Milky Way, we assume $M_{\rm{MW}}=
6.43\times10^{10}~\rm{M_{\odot}}$ (McMillan 2011). We note that
although the GC metallicities for the Milky Way and early-type
galaxies are derived from different methods (i.e. CMDs vs CaT
spectra), systematic differences in the metallicity scale are less
than 0.2 dex (Usher et al. 2012).

The mean MPGC and MRGC metallicities and stellar mass for the 11
early-type galaxies and the Milky Way are given in Table 1. We include
NGC 4494 but note that it appears to have a trimodal rather than
bimodal metallicity distribution (Usher et al. 2012) which may explain
its very metal-rich MRGC mean metallicity.  Our early-type galaxy
sample has a mean stellar mass of log M$_{\ast}$ = 11.2 with mean
metallicities [Z/H] of --1.23 $\pm$ 0.03 and --0.39 $\pm$ 0.02 for the
MPGC and MRGC subpopulations respectively. Thus the MPGC mean metallicity is comparable to 
that of the Milky Way, but the MRGC subpopulation is somewhat more metal-poor than that of 
the Milky Way.

\begin{table}
\caption{The early-type galaxy sample and the Milky Way. Galaxy name, mean metallicity of the metal-poor and metal-rich globular 
clusters with the number of globular clusters in each subpopulation, 
and galaxy stellar mass. }
\begin{tabular}{lccccc}

\hline
Galaxy & MPGC [Z/H] & N & MRGC [Z/H]  & N & M$_{\ast}$\\
 (NGC)      & (dex)  & & (dex) & & (M$_{\odot}$)\\
\hline

1023	& $-1.34\pm0.10$ & $8$  & $-0.47\pm0.12$ & $10$	& $8.24\times10^{10}$ \\
1400	& $-1.25\pm0.19$ & $13$ & $-0.50\pm0.18$ & $11$	& $1.08\times10^{11}$ \\
1407	& $-1.23\pm0.08$ & $70$	& $-0.39\pm0.05$ & $79$	& $3.13\times10^{11}$ \\
2768	& $-1.48\pm0.14$ & $14$ & $-0.68\pm0.11$ & $19$	& $1.57\times10^{11}$ \\
3115	& $-1.27\pm0.06$ & $41$	& $-0.14\pm0.06$ & $47$	& $8.17\times10^{10}$ \\
4278	& $-1.45\pm0.07$ & $46$	& $-0.63\pm0.05$ & $77$	& $6.79\times10^{10}$ \\
4365	& $-1.39\pm0.12$ & $29$ & $-0.40\pm0.05$ & $71$ & $2.49\times10^{11}$ \\
4473	& $-1.14\pm0.07$ & $35$	& $-0.12\pm0.13$ & $11$	& $6.61\times10^{10}$ \\
4494	& $-1.16\pm0.10$ & $23$ & $+0.06\pm0.11$ & $15$ & $9.12\times10^{10}$ \\
4649	& $-1.04\pm0.06$ & $71$	& $-0.35\pm0.07$ & $43$	& $2.86\times10^{11}$ \\
5846	& $-1.05\pm0.16$ & $19$	& $-0.37\pm0.12$ & $16$	& $2.05\times10^{11}$ \\
\hline
MW  		& $-1.24\pm0.04$ & $109$ & $-0.24\pm0.03$ & $43$ & $6.43\times10^{10}$ \\

\hline

\end{tabular}
\end{table}

\section{An Evolving Mass-Stellar Metallicity Relation}

		In order to define the redshift evolution of the
		galaxy mass-metallicity relation we follow the
		approach of Kruijssen (2014). This involves creating a
		functional form for an evolving mass-metallicity relation
		based on observations of the {\it gas} metallicity at
		z $\sim$ 0.1 (Tremonti et al. 2004), z $\sim$ 2 (Erb
		et al. 2006) and z $\sim$ 3--4 (Mannucci et al. 2009).
We expect the evolving mass-metallicity relation to be well defined up to redshift z = 4 in the coming years with surveys 
such as MOSDEF (Sanders et al. 2015). 

Although there is a wide range of methods for studying gas metallicity in the literature, when a given method is 
		used the mass-metallicity relations generally agree within 0.1 dex (Kewley \& Ellison 2008). 
 		Gas metallicity traces the current chemical enrichment of a system, whereas the stellar metallicity is an average value over its
star formation history. For GCs, the mean stellar metallicity is imprinted at birth and doesn't change with time.
So following Shapiro et al. (2010) we further assume that the gas metallicity is 
equal to the total stellar metallicity, i.e. [O/H] $\approx$ [Z/H].	

		In particular, we adopt 
		equation 5 of Kruijssen (2014):
		\begin{eqnarray}
			[Z/H] \approx [O/H] = -0.59+0.24\log\frac{M_{\rm{\star}}}{10^{10}}   \nonumber \\
 -8.03\times10^{-2}\times\left[\log \frac{M_{\rm{\star}}}{10^{10}}\right]^2  	-0.2(z-2)+0.94[\alpha/Fe]
		\end{eqnarray}
		where $z$ is the redshift, $M_{\rm{\star}}$ is the total stellar 
		mass expressed in $\rm{M_{\odot}}$ and the $0.94[\alpha/Fe]$ term converts 
[Fe/H] to [Z/H] using Eq. 1. 
We adopt a mean value for $[\alpha/Fe]$ of 0.3 (Pritzl et al. 2005) and note that 
there is little or no systematic dependence of the ratio on metallicity, i.e. MPGCs and MRGCs have similar $[\alpha/Fe]$ ratios. 
We also note that the galaxy mass-metallicity relation shows little or no variation with environment up to z = 2 (Kacprzak et al. 2015) and that it has the same slope 
up to a redshift of at least 6 according to a recent simulation by Ma et al. (2015). 
As individual GCs are essentially single stellar populations (at least compared to the complex stellar 
populations of their host galaxies) we can use measurements of their mean (stellar) 
metallicity to estimate their epoch of formation given knowledge 
of their host galaxy stellar mass.


\section{Results}
 
 \begin{figure*}
\begin{center}
\includegraphics[scale=0.7,angle=0]{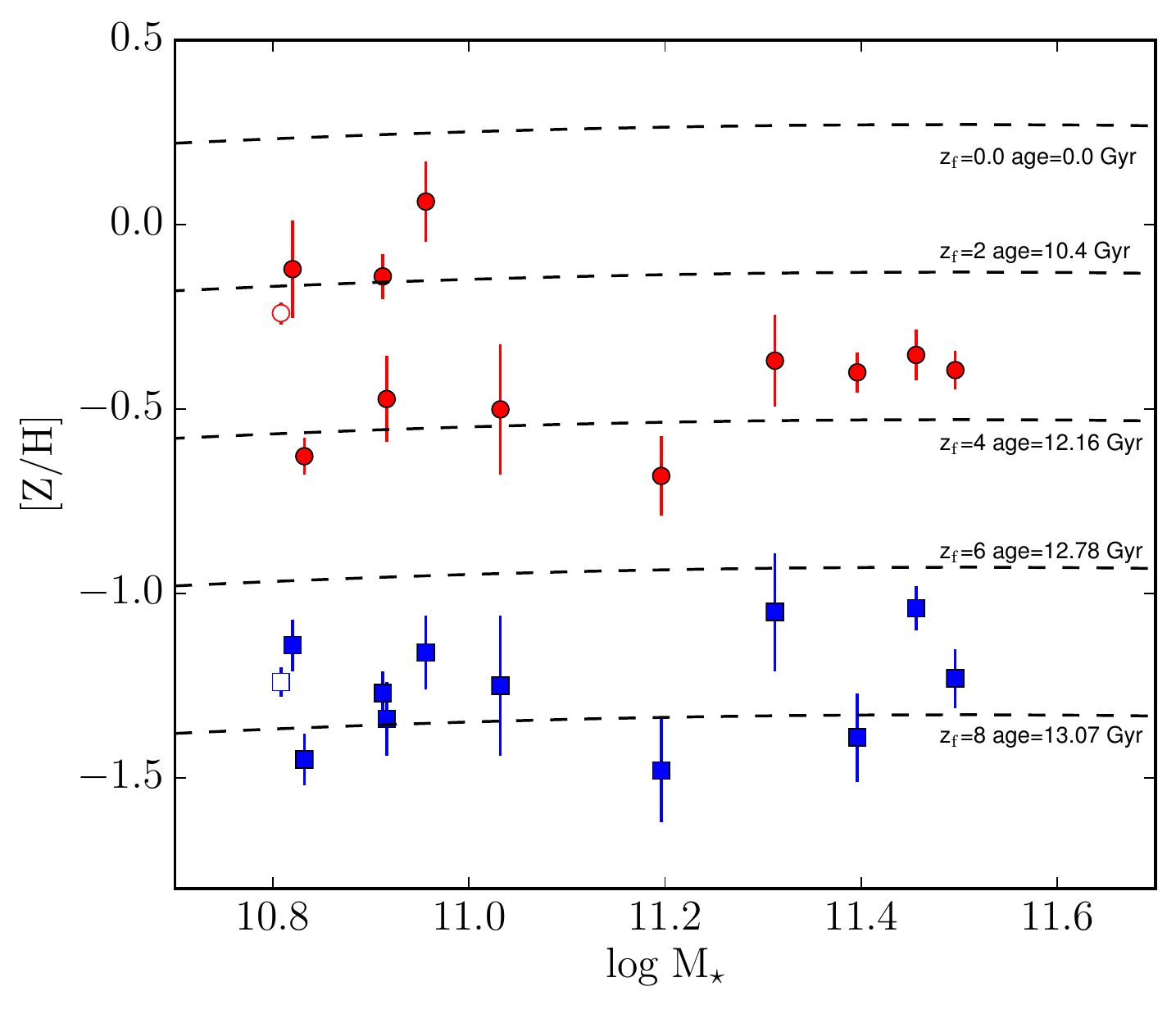}
\caption{Evolving mass-metallicity relations. 
	    		The solid red circles and blue squares show, 
respectively, the average metallicities 
                        for the MRGC and MPGC subpopulations in 11 early-type 
                        galaxies from the SLUGGS survey. The error bars represent the error on the mean. The open red and blue points 
                        represent the MRGCs and MPGCs for the Milky Way. 
	    		The dashed black lines show the evolving mass-metallicity relations using  
Eq. 2 (see text for details), and are labelled with the associated formation redshift and look-back age. 
}
\end{center}
\end{figure*}
 		
In Figure 1 we show the mean metallicity (and error on the mean) for the individual 
MPGC and MRGC subpopulations of  our 
early-type galaxy sample and the Milky Way as a function of host galaxy stellar mass. 
The galaxy mass-stellar metallicity relation at various redshifts (Eq. 2) is also shown.

The MRGCs  
reveal more scatter (and have larger error bars) at low galaxy masses towards younger mean formation ages. This scatter, even at a fixed galaxy mass, may imply a range of MRGC formation histories. 
The Milky Way GC system is well-known to host 
several intermediate-age metal-rich GCs that may be associated with accreted
satellite galaxies (Forbes \& Bridges 2010); a similar situation may be present in low mass early-type galaxies. 
We note that the stellar 
mass plotted for the Milky Way is the total mass, whereas MRGCs are mostly associated 
with the bulge which has a much lower 
stellar mass of $\sim10^{10}$ M$_{\odot}$. 
The MPGCs in our sample reveal less scatter 
in their mean metallicities than the MRGCs, suggesting more uniformly old formation ages. 
For their {\it current} host galaxy stellar masses, the  MRGCs and MPGCs in the early-type 
galaxies have best fit average  
formation epochs of z = 3.4 and z = 7.4 respectively. Similarly, the MRGC and MPGC 
subpopulations of the Milky Way correspond 
to z = 2.4 and z = 7.4, if the GCs formed in the Milky Way with 
its current stellar mass. 

These formation epochs assume that both
GC subpopulations formed within the host galaxy with the mass 
that it has today (i.e. the GCs formed in-situ within 
the host galaxy, and the galaxy has not grown in stellar mass after the GCs formed). However, studies of compact, quiescent galaxies at high redshift have concluded
that they have grown in mass by a factor of several due to the
merger/accretion of lower mass galaxies (Bezanson et al. 2009; Marchesini et al. 2014; Ownsworth et al. 2014; Stefanon et al. 2015).
Thus if GCs have formed in-situ, we must consider the mass of the host galaxy at the time of GC formation. Furthermore, GCs may have formed ex-situ, i.e. within a low mass satellite that has 
been accreted into the halo of the main host galaxy. We consider both the in-situ and accreted 
origins for GCs in thie discussion below.


Similarities between galaxy stellar properties and MRGCs, e.g. radial density profiles and azimuthal distributions (e.g. Kartha et al. 2014), 
suggest that a significant fraction of MRGCs formed
in-situ within the host galaxy where they now reside.  
The formation site for MPGCs is less clear. They have more 
extended radial density profiles than the galaxy starlight indicating a significant 
contribution from accretion. However, their inner 
radial metallicity gradients are similar to those of the MRGCs 
(Harris 2009; Forbes et al. 2011; Pastorello et al. 2015), suggesting that 
some MPGCs may have also formed in-situ within the same host galaxy as the MRGCs. 
 

Accretion of low mass satellites, and their 
GCs, is thought to be particularly important in the two-phase
formation of massive galaxies (e.g. Oser et al. 2012). The accretion
process is expected to be fairly self-similar, with a typical accreted
satellite mass having a mass some 10 per cent of its host galaxy (Stewart et
al. 2008; Oser et al. 2012; Hirschmann et al. 2015). For our sample of early-type galaxies 
with a mean mass of log M$_{\ast}$ = 11.2, this implies that the accreted satellite mass 
could be as high as log M$_{\ast}$ =
10.2 M$_{\odot}$ (similar to the mass of M33).  
We do not expect many MRGCs to be accreted since galaxies with log M$_{\ast}$
$<$ 10.2 are increasingly unimodal, and dominated by MPGCs (Forbes 2005).

For the Milky Way a satellite
galaxy of mass a few 10$^9$ M$_{\odot}$ would be more typical. 
This is similar to the mass of the Sgr dwarf 
(Niederste-Ostholt et al. 2010), and the recent findings of D'Abrusco
et al. (2015) for accreted substructures in GC systems.
We note that surviving dwarf galaxies below this mass host very few
GCs, e.g. the WLM galaxy with a baryonic mass mass of 10$^8$  
M$_{\odot}$ (Leaman et al. 2012) hosts only a single metal-poor GC.

In Figure 2 we illustrate the inferred formation epochs for an in-situ or accreted origin for the 
GCs in our early-type galaxy sample. 
The two red and blue curved bands show the formation epochs of
the MRGC and MPGC subpopulations, derived from the best fit of their
mean metallicities to the mass-metallicity relation (Eq. 2), while
allowing the host galaxy mass to be a fraction of today's mass. 
The shaded regions represent the error
on the mean and a possible systematic uncertainty of $\pm$0.2 dex.

We show the predicted stellar growth of a galaxy with log M$_{\ast}$ $\sim$ 11.2 
(i.e. the mean mass of our sample), and a galaxy with 10 per cent of that mass today, 
to represent 
a typical accreted satellite for our sample of massive early-type galaxies. We label these 
curves `in-situ' and `accreted' respectively to represent the stellar mass growth of the main host galaxy 
and a typical satellite with look-back time. These curves come from 
Behroozi et al. (2013), who compared the merger trees of simulated dark matter halos 
from z = 8 to z = 0  
with observations of the star formation efficiency and the stellar mass function.

The in-situ growth line intersects the MRGC curve at log M$_{\ast}$
$\sim$ 10.6, suggesting a host galaxy mass for MRGC formation some 4
times less than today. It corresponds to a formation epoch of z = 2.9
or a look-back age of 11.5 Gyr. We note that if instead of the
Behroozi et al. relation, we had adjusted the observed linear relation of
Ownsworth et al. (2014) from their stellar mass of log M$_{\ast}$ =
11.56 to match our stellar mass, and used their stellar mass
growth from redshifts z = 0.3 to z = 3, it would result in only a
small shift in the inferred formation redshift of MRGCs to z $\sim$
2.8. This gives us some confidence in the robustness of our results.
MPGCs that form in-situ within the main host galaxy had a mean
formation epoch of z = 5.9 (12.8 Gyr) and formed in a galaxy 1/50th of
today's mass.

For MPGCs that were accreted, the
accreted satellite growth line intersects the MPGC curve at log
M$_{\ast}$ $\sim$ 8.8, suggesting that a typical satellite was 1/25th
of today's mass when MPGCs formed. The corresponding formation
epoch for these accreted MPGCs is 4.8 (12.5 Gyr). 
Figure 2 suggests that any accreted MRGCs would have a mean formation
epoch of z = 2.3 (10.8 Gyr), but we expect such MRGCs to be a minor 
contribution to the in-situ formed MRGCs.

Figure 2 also shows a shaded region representing the epoch of hydrogen
reionisation (Robertson et al. 2015). 
Several GC formation models invoke reionisation as the
mechanism to terminate the formation of MPGCs and hence provide a
metallicity and temporal gap before MRGC formation occurs. Although
the onset of reionisation is still poorly defined, the epoch of
instantaneous reionisation inferred from Planck data is z = 8.8 and a
variety of different observations indicate that reionisation had
essentially ended by z = 6 (see Robertson et al. 2015 for a
summary). The figure shows that for our early-type galaxy sample, only
MPGCs that form in host galaxies with mass log M$_{\ast}$ $\ge$ 9.5 
are affected by reionisation.  Below this galaxy mass, a process
other than reionisation is needed to establish the distinct mean
metallicities observed in MPGC and MRGC subpopulations. A dependence
of the typical accreted satellite mass with redshift (e.g. Muratov \&
Gnedin 2010) is one possible mechanism. 

\begin{figure*}
\begin{center}
\includegraphics[scale=0.7,angle=0]{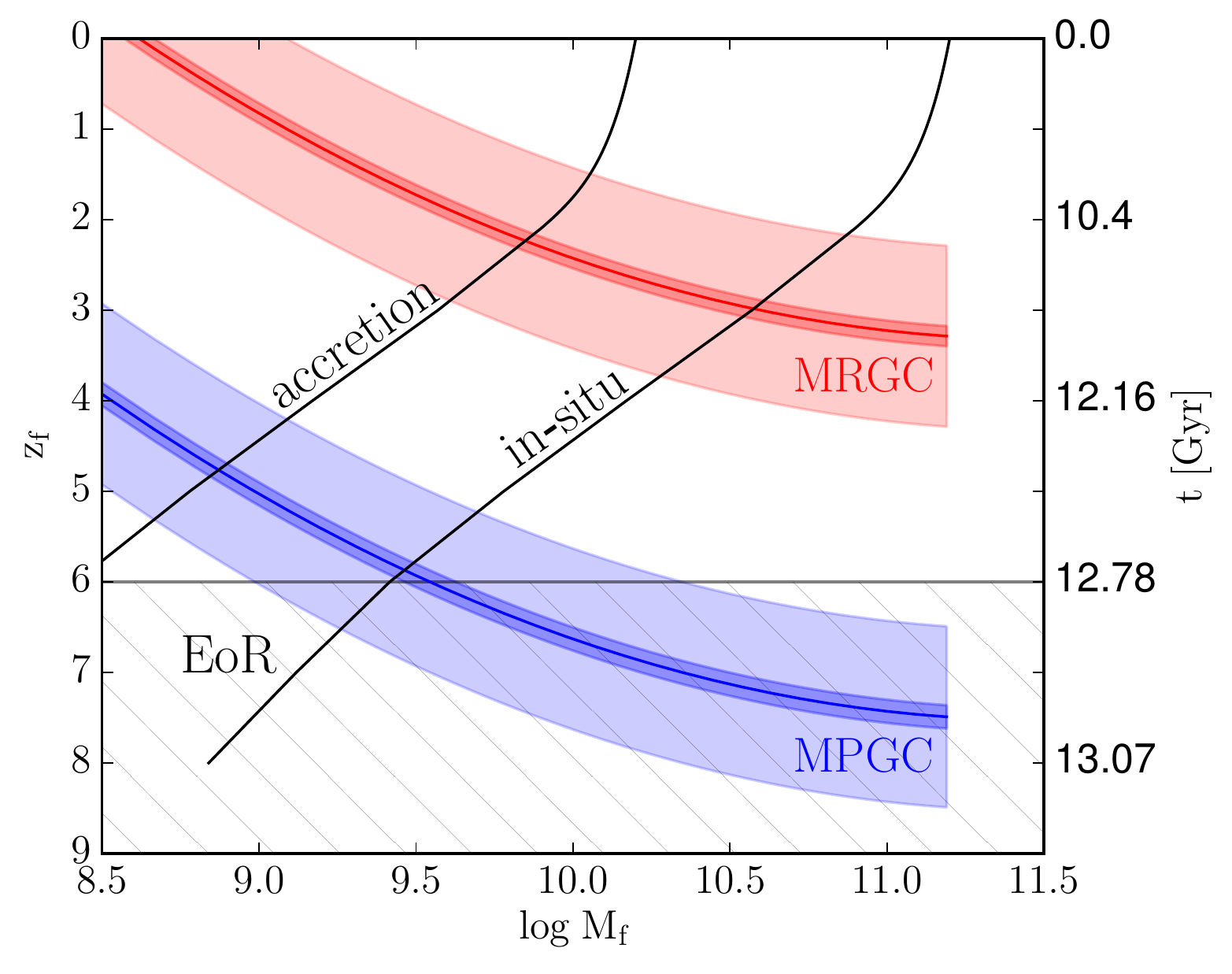}
\caption{
Formation epoch for MPGCs and MRGCs. The formation epoch is shown as a
function of galaxy stellar mass at formation. The blue and red curved bands
show the MPGC and MRGC formation epochs for a range of galaxy
formation mass ranging from a host galaxy with the mean stellar mass
of our sample (i.e. log M$_{\ast}$ = 11.2). The dark shaded regions
show the error on the mean metallicity, and the light shaded regions
show the possible range of systematic uncertainty in the mean
metallicity. The grey hashed region at the bottom of the plot shows
the epoch of reionisation (EoR).  The curved black solid lines show
the predicted stellar mass growth from z = 8 by Behroozi et al. (2013). The `in-situ' 
curve shows the growth of a log M$_{\ast}$ = 11.2 galaxy today, while the `accreted' 
curve shows the growth of a satellite of mass 10 per cent that of our sample, i.e. 
log M$_{\ast}$ = 10.2. 
In the simplified picture that MRGC form in-situ within the main host galaxy, and 
MPGCs are accreted from satellites, with implied formation epochs of z = 
2.9 (11.5 Gyr) and 
z = 4.8 (12.5 Gyr) respectively. 
In reality, MPGCs may have a significant in-situ formed component, and 
accretion may contribute some MRGCs, thus increasing the inferred age difference between the two 
GC subpopulations.  
}
\end{center}
\end{figure*}

We have performed a similar analysis for the Milky Way GC
system. Using the mean metallicity of the Milky Way's MRGC and MPGCs (Table 1)
and assuming that a Behroozi et al. (2013) mass growth history for a 
stellar mass log M$_{\ast}$ = 10.8 today is applicable 
to the Milky Way, we find formation epochs of z = 2.2 (10.7 Gyr) 
and 5.5 (12.7 Gyr) for in-situ formed MRGCs and MPGCs respectively. 
For MRGCs and MPGCs accreted from satellites of mass 10 per cent of the Milky Way, the 
formation epochs are z = 1.8 (10 Gyr) and z = 4.1 (12.2 Gyr) respectively.

A comparison of our findings with cosmological predictions
(Section 2) and observations of the Milky Way's GC system (Section 3) are 
given in Table 2, 
in which we list the mean look-back ages (or a range if mean ages are not available)  
and corresponding 
redshifts. Our findings are discussed in the next section.

Our mean ages are potentially affected by a number of systematic
effects, which dominate over any error on the mean value. These
include the absolute [O/H] metallicity scale, bias in our
spectroscopic metallicities, variations in the colour to metallicity
conversion, and instrinsic scatter in $[\alpha/Fe]$ ratios. We
conservatively estimate the combined systematic uncertainty to be less
than $\pm$0.2 dex. 
The corresponding uncertainty in the formation epoch from
the systematic effects results in a redshift uncertainty of $\pm$ 1. 
In terms of age, the systematic uncertainty is up to $+0.6$,
$-1.2$ Gyr for the MRGCs which have a mean age of 11.5 Gyr.  For the
MPGCs, the systematic uncertainty is $+0.2$, $-0.3$ for in-situ formed
MPGCs with a mean age of 12.8 Gyr, and $+0.2$, $-0.4$ for MPGCs formed
in satellites with a mean age of 12.5 Gyr.  As well as increasing the
sample size, future efforts to apply this method should be directed at
reducing possible systematic effects.


\begin{table}
\caption{Age and redshift of formation for GC subpopulations.  
Predictions from cosmological models, 
Milky Way observations and the results from this work on 11 early-type 
galaxies and the Milky Way. We assume a Bennett et al. (2014) cosmology.}
\begin{tabular}{lccc}

\hline
Author & MPGC & MRGC & $\Delta$Age\\
       & Gyr (z) & Gyr (z) & Gyr\\
\hline
Beasley & 12.7 (5.8) & 10.2 (1.9) & 2.5\\
Santos & $>$13.0 ($>$7) & $>$11.5 ($>$3) & $<$1.5\\
Bekki & 12.7 (5.7)& 12.3 (4.3) & 0.4\\
Griffen & 13.4-13.6 (13-22) & 7-13.3 (0.8-11)& 0.1-6.6\\
Tonini & 11.5-12.2 (3-4) & 10.4 (2) & 1.1-1.8\\
Li & 12.2-12.8 (4-6) & 7.8-12.2 (1-4) & 0-5\\
Katz$^{\dagger}$ & 13.0-13.4 (7-12) & 10.4 (2) & 2.6-3.0\\
Corbett & 13.1 (9) & -- & --\\
Trenti & 13.2 (9.3) & -- & --\\
\hline
MW ages & 12.5-12.8 (5-6.1 )& 11.0-12.8 (2.4-6.1)& 0-1.5\\
\hline
11 ETGs & 12.5-12.8 (4.8-5.9) & 11.5 (2.9) & 1.0-1.3\\
MW & 12.2-12.7 (4.1-5.5) & 10.7 (2.2) & 1.5-2.0\\
\hline

\end{tabular}
$^{\dagger}$ predictions for GCs in Milky Way type galaxies. 

\end{table} 

\section{Discussion and Conclusions}

The mass-metallicity relation for massive galaxies is a strong function of redshift and only a weak function of galaxy mass. Here we exploit this 
fact, and new measurements of the mean metallicity in 11 massive early-type galaxies from the SLUGGS survey, to  estimate the 
mean formation epoch of metal-rich (MR) and metal-poor (MP) globular clusters (GCs). 
We find more scatter in the mean metallicity of the MRGCs  than for the  
MPGCs. 

We infer a
formation epoch for MRGCs of z = 2.9 and hence a mean age of 11.5
Gyr. We expect the contribution of (slightly younger) 
MRGCs formed in satellites and later accreted into the 
halos of early-type galaxies to be small. 
Our MRGC mean age is broadly in line with most cosmological model predictions
for GC formation. 
Although we find a younger age than the mean of 12.3
Gyr predicted by Bekki et al. (2008), their simulations suggested that
a wide range of MRGC ages are possible.
Using a similar method for the Milky Way we find a mean age of 
10.7 Gyr for the MRGCs. This is somewhat younger than the MRGCs associated with the 
bulge of the Milky Way.

We infer mean ages for the MPGCs to be 12.5 Gyr if they all accreted from satellites and 12.8 Gyr if they are all formed in-situ within the main host galaxy. 
Our MPGC mean age limits of 12.5--12.9 Gyr are consistent with 
predictions from several cosmological models. 
However, our results disfavour models that predict very early MPGC 
formation (ages of $\ge$ 13.2 Gyr and z $\ge$ 9) in dark matter minihalos.
For the MPGCs of the Milky Way we infer mean age limits 
of between 12.2--12.7 Gyr, which is 
consistent with current observational results. The MPGC mean ages from
our work, and Milky Way observations, suggest that MPGCs continued to
form long after the end of reionisation, casting doubt on simulations that
invoke reionisation to explain the observed metallicity differences
between the two GC subpopulations, or indeed, invoke GCs as the main source 
of reionisation. 

We note that the only cosmological model that is consistent with {\it
both} our MPGC and MRGC mean ages is that of Li \& Gnedin
(2014). 
In their model the vast bulk of GCs form after 
reionisation has ended, and metallicity bimodality is 
established by MPGCs that form at high redshift, i.e. z $\sim$ 4--6 
(12.2--12.8 Gyr)   
and MRGCs which peak around z $\sim$ 3--4 (11.5--12.2 Gyr)  
but can continue to form to z $\sim$ 1 (7.8 Gyr).

Historically it has been very difficult to verify directly an age
difference in extragalactic GC subpopulations, and hence whether they
formed in independent formation episodes. Here we infer a mean age
difference between the MPGC and MRGC subpopulations in masive early-type 
galaxies of between 1.0 and
1.3 Gyr. This age difference is somewhat smaller than the mean
difference predicted by Beasley et al. (2002) and larger than that of Bekki 
et al. (2008). 
For the Milky Way we infer a difference of 1.5--2 Gyr. The prediction from 
the  Katz \& Ricotti (2014) simulation of Milky Way like galaxies is 
2.6--3.0 Gyr. An age difference of 1.5 Gyr was measured for the Milky Way's
GC system by VandenBergh et al. (2013), while no measurable age
difference was found by Marin-Franch et al. (2009). 

Future high resolution hydrodynamical simulations in the relevant
redshift range of 2 $<$ z $<$ 7 could reveal important clues regarding the
relative roles of in-situ and ex-situ (accreted) GCs.
Additional work is also needed on the observational side to better
understand systematic uncertainties in applying an evolving
mass-metallicity relation to determine the formation epoch of
GCs. Nevertheless larger samples of GCs over a range of host galaxy
masses offer the potential for interesting constraints on the mean age
of GCs and galaxy assembly in general.


\section{Acknowledgements}

We thank L. Cortese and L. Spitler for useful discussions. 
DAF thanks the ARC for financial support via DP130100388. 
This work was supported by NSF grant AST-1211995. Finally, we thank the 
referee for a prompt report 
and several useful suggestions to improve the paper.

\section{References}

Bailin J., Harris W.~E., 2009, ApJ, 695, 1082 \\
Baistan N., Cabrera-Ziri I., Davies B., Larsen S.~S., 2013, MNRAS, 436, 2852 \\
Beasley M.~A., Baugh C.~M., Forbes D.~A., 
Sharples R.~M., Frenk C.~S., 2002, MNRAS, 333, 383 \\
Bedin L.~R., Salaris M., Piotto G., Anderson J., King I.~R., Cassisi S., 
2009, ApJ, 697, 965 \\
Behroozi P.~S., Wechsler R.~H., Conroy C., 2013, ApJ, 770, 57 \\
Bekki K., Yahagi H., Nagashima M., Forbes D.~A., 2008, MNRAS, 387, 1131 \\
Bennett  C.~L., Larson D., Weiland J.~L., 
Hinshaw G., 2014, ApJ, 794, 135 \\
Bezanson R., van Dokkum P.~G., Tal T., 
Marchesini D., Kriek M., Franx M., Coppi P., 2009, ApJ, 697, 1290 \\
Boley A.~C., Lake G., Read J., Teyssier R., 2009, ApJ, 706, L192 \\
Brodie J.~P., Strader J., 2006, ARA\&A, 44, 193 \\
Brodie J.~P., et al., 2014, ApJ, 796, 52\\
Bromm V., Clarke C.~J., 2002, ApJ, 566, L1 \\
Corbett Moran C., Teyssier R., Lake G., 2014, MNRAS, 442, 2826 \\
D'Abrusco R., Fabbiano G., Zezas A., 2015, ApJ, 805, 26\\
Erb D.~K., Shapley A.~E., Pettini M., Steidel C.~C., Reddy N.~A., 
Adelberger K.~L., 2006, ApJ, 644, 813 \\
Forbes D.~A., 2005, ApJ, 635, L137 \\
Forbes D.~A., Bridges T., 2010, MNRAS, 404, 1203 \\
Forbes D.~A., Spitler L.~R., Strader J., Romanowsky A.~J., Brodie J.~P., 
Foster C., 2011, MNRAS, 413, 2943 \\
Garcia-Berro E., Torres S., Althaus L.~G., Miller Bertolami M.~M., 2014, A\&A, 571, A56 \\
Griffen B.~F., Drinkwater M.~J., Thomas \\
P.~A., Helly J.~C., Pimbblet K.~A., 2010, MNRAS, 405, 375 \\
Hansen B.~M.~S., et al., 2013, Natur, 500, 51\\
Harris W.~E., 1996, AJ, 112, 1487 \\
Harris W.~E., 2009, ApJ, 703, 939 \\
Hirschmann M., Naab T., Ostriker J.~P., 
Forbes D.~A., Duc P.-A., Dav{\'e} R., Oser L., Karabal E., 2015, MNRAS, 
449, 528 \\
Jarrett T.~H., Chester T., Cutri R., 
Schneider S., Skrutskie M., Huchra J.~P., 2000, AJ, 119, 2498 \\
Kacprzak G.~G., et al., 2015, ApJ, 802, L26 \\
Kartha S.~S., Forbes D.~A., Spitler L.~R., Romanowsky A.~J., Arnold J.~A., 
Brodie J.~P., 2014, MNRAS, 437, 273 \\
Katz H., Ricotti M., 2014, MNRAS, 444, 2377 \\
Kewley L.~J., Ellison S.~L., 2008, ApJ, 681, 1183 \\
Kruijssen J.~M.~D., 2014, CQGra, 31, 244006 \\
Lada C.~J., Lada E.~A., 2003, ARA\&A, 41, 57\\
Leaman R., et al., 2012, ApJ, 750, 33\\
Li H., Gnedin O.~Y., 2014, ApJ, 796, 10 \\
Ma X., et al. 2015, arXiV:1504.02097\\
Mannucci F., et al., 2009, MNRAS, 398,  1915 \\
Marchesini D., et al., 2014, ApJ, 794, 65 \\
Marin-Franch A., et al., 2009, ApJ, 
694, 1498 \\
McMillan P.~J., 2011, MNRAS, 414, 2446 \\
Muratove A.~L., Gnedin O.~Y., 2010, ApJ, 718, 1266 \\
Pastorello N., Forbes D.~A., Foster C., 
Brodie J.~P., Usher C., Romanowsky A.~J., Strader J., Arnold J.~A., 2014, 
MNRAS, 442, 1003 \\
Niederste-Ostholt M., Belokurov V., Evans 
N.~W., Koposov S., Gieles M., Irwin M.~J., 2010, MNRAS, 408, L66 \\
Ownsworth J.~R., Conselice C.~J., Mortlock 
A., Hartley W.~G., Almaini O., Duncan K., Mundy C.~J., 2014, MNRAS, 445, 
2198 \\
Oser L., Naab T., Ostriker J.~P., Johansson P.~H., 2012, ApJ, 744, 63 \\
Pastorello N., et al. 2015, arXiv:1505.04795\\ 
Peng E.~W., et al., 2006, ApJ, 639, 838\\
Piotto G., et al., 2007, ApJ, 661, L53 \\
Pritzl B.~J., Venn K.~A., Irwin M., 2005, AJ, 130, 2140 \\
Puzia T.~H., Kissler-Patig M., Thomas D., Maraston C., Saglia R.~P., Bender R., Goudfrooij P., Hempel M., 2005, A\&A, 439, 997 \\
Robertson B.~E., Ellis R.~S., Furlanetto 
S.~R., Dunlop J.~S., 2015, ApJ, 802, L19 \\
Sanders R.~L., et al., 2015, ApJ, 799, 138 \\
Santos  M.~R., 2003, egcs.conf, 348 \\
Shapiro K.~L., Genzel R., F{\"o}rster Schreiber N.~M., 2010, MNRAS, 403, L36 \\
Spitler L.~R., 2010, MNRAS, 406, 1125 \\
Spitler L.~R., Romanowsky A.~J., Diemand 
J., Strader J., Forbes D.~A., Moore B., Brodie J.~P., 2012, MNRAS, 423, 
2177 \\
Springel V., et al., 2005, Natur, 435, 629 \\
Strader J., Brodie J.~P., Cenarro A.~J., 
Beasley M.~A., Forbes D.~A., 2005, AJ, 130, 1315\\
Strader J., Smith G.~H., 2008, AJ, 136, 1828 \\
Stefanon M., et al., 2015, ApJ, 803, 11 \\
Stewart K.~R., Bullock J.~S., Wechsler 
R.~H., Maller A.~H., Zentner A.~R., 2008, ApJ, 683, 597 \\
Thomas D., Maraston C., Bender R., 2003, MNRAS, 339, 897\\
Tonini C., 2013, ApJ, 762, 39 \\
Tremonti C.~A., et al., 2004, ApJ, 613,  898 \\
Trenti M., Perna R., Jimenez R., 2015, ApJ, 802, 103 \\
Usher C., et al., 2012, MNRAS, 426, 1475\\
Usher C., et al., 2015, MNRAS, 446, 369 \\
VandenBerg D.~A., Brogaard K., Leaman R., 
Casagrande L., 2013, ApJ, 775, 134 \\
Vazdekis A., Cenarro A.~J., Gorgas J., 
Cardiel N., Peletier R.~F., 2003, MNRAS, 340, 1317 \\
Wright  E.~L., 2006, PASP, 118, 1711 \\

\end{document}